\documentclass[aps,prl,twocolumn,superscriptaddress,nobibnotes,letterpaper]{revtex4}
\usepackage{times}
\usepackage[pdftex]{graphicx}
\usepackage[pdftex]{epsfig}
\usepackage{amsmath}
\usepackage{pslatex}
\usepackage{amssymb}
\usepackage{amsfonts}
\usepackage{color}
\usepackage{wrapfig}
\usepackage{eucal}
\usepackage{hhline}
\usepackage{threeparttable}
\usepackage{supertabular}
\usepackage{multirow}
\usepackage{tabularx}
\usepackage{setspace}
\usepackage[compact]{titlesec}
\usepackage{hyperref}
\usepackage{graphicx}
\usepackage{esint}
\usepackage{verbatim}
\usepackage{units}

\newcommand{\gzeroO}{g_{\text{0,o}}}
\newcommand{\gOMO}{g_{\text{OM}}}

\newcommand{\omegacO}{\omega_{\text{c}}}
\newcommand{\lambdacO}{\lambda_{\text{c}}}
\newcommand{\omegamO}{\omega_{\text{m}}}

\newcommand{\QoO}{Q_{\text{c}}}
\newcommand{\QmO}{Q_{\text{m}}}

\newcommand{\mO}{m_{\text{eff}}}
\newcommand{\kO}{k_{\text{eff}}}

\newcommand{\dUe}{\delta \bar{U}_{\text{E}}}
\newcommand{\Ues}{U_{\text{es}}}

\newcommand{\gzeroE}{g_{\text{0,e}}}

\newcommand{\omegacE}{\omega_{\text{LC}}}

\newcommand{\etaE}{\eta_{\text{e}}}

\newcommand{\GLE}{G_{\text{e,1}}}
\newcommand{\GNLE}{G_{\text{e,2}}}
\newcommand{\xzpfE}{x_{\text{zpf}}}

\newcommand{\VDC}{V_{\text{DC}}}
\newcommand{\VAC}{V_{\text{AC}}}

\newcommand{\Fcap}{F_{\text{cap}}}

\newcommand{\Cs}{C_{\text{s}}}
\newcommand{\Cm}{C_{\text{m}}}
\newcommand{\gOME}{g_{\text{EM}}}
\newcommand{\vecb}[1]{\mathbf{#1}}
\newcommand{\Vrms}{V$_{\text{rms}}$}

\begin{document}

\title{Linear and nonlinear capacitive coupling of electro-opto-mechanical photonic crystal cavities}

\author{Alessandro Pitanti}
\affiliation{Institute for Quantum Information and Matter and Thomas J. Watson, Sr., Laboratory of Applied Physics, California Institute of Technology, Pasadena, CA 91125, USA}
\affiliation{NEST and Istituto Nanoscienze - CNR, Scuola Normale Superiore, Piazza San Silvestro 12, 56127 Pisa, Italy}
\author{Johannes M. Fink}
\affiliation{Institute for Quantum Information and Matter and Thomas J. Watson, Sr., Laboratory of Applied Physics, California Institute of Technology, Pasadena, CA 91125, USA}
\author{Amir H. Safavi-Naeini}
\author{Chan U. Lei}
\author{Jeff T. Hill}
\affiliation{Institute for Quantum Information and Matter and Thomas J. Watson, Sr., Laboratory of Applied Physics, California Institute of Technology, Pasadena, CA 91125, USA}
\author{Alessandro Tredicucci}
\affiliation{NEST and Istituto Nanoscienze - CNR, Scuola Normale Superiore, Piazza San Silvestro 12, 56127 Pisa, Italy}
\affiliation{3 Dipartimento di Fisica, Università di Pisa, Largo Pontecorvo 3, 56127 Pisa, Italy}
\author{Oskar Painter}
\email{opainter@caltech.edu}
\affiliation{Institute for Quantum Information and Matter and Thomas J. Watson, Sr., Laboratory of Applied Physics, California Institute of Technology, Pasadena, CA 91125, USA}

\date{\today}
\begin{abstract}
We fabricate and characterize a microscale silicon electro-opto-mechanical system whose mechanical motion is coupled capacitively to an electrical circuit and optically via radiation pressure to a photonic crystal cavity. To achieve large electromechanical interaction strength, we implement an inverse shadow mask fabrication scheme which obtains capacitor gaps as small as $30$~nm while maintaining a silicon surface quality necessary for minimizing optical loss. Using the sensitive optical read-out of the photonic crystal cavity, we characterize the linear and nonlinear capacitive coupling to the fundamental $\omegamO/2\pi = 63$~MHz in-plane flexural motion of the structure, showing that the large electromechanical coupling in such devices may be suitable for realizing efficient microwave-to-optical signal conversion.
\end{abstract}
\maketitle

Microelectromechanical systems (MEMS) are a widespread technology platform with a vast number of applications.  MEMS devices are found, for instance, in a variety of hand-held electronic devices, often as accelerometers~\cite{Yazdi1998}, microphones~\cite{Tajima2003} or pressure sensors~\cite{Eatony1997}. Recently, MEMS have been proposed for energy harvesting applications~\cite{Cook-Chennault2008}, ultra-high resolution mass sensors~\cite{Hanay2012} and as a suitable candidate for the development of biological sensors in “lab-on-a-chip” technologies~\cite{Manz1990}. New generation of MEMS have critical dimensions down below the microscale, and into the nanoscale, opening the possibility to integrate these devices with other nanotechnologies.  In the case of nanophotonics, the emerging field of cavity optomechanics uses the radiation pressure force to probe and control the state of a mechanical actuator embedded in an optical cavity. Optomechanical nanophotonic devices, such as photonic crystal ``zipper'' cavities~\cite{Eichenfield2009a} and optomechanical crystals (OMCs)~\cite{Eichenfield2009}, have been proven effective for near quantum-limited position read-out~\cite{Cohen2013,Anetsberger2010} and strong back-action effects, as shown by the cooling of a mechanical mode to its quantum ground state of motion~\cite{Chan2011}. 

The integration of MEMS with optomechanical devices may be useful both for microphotonic circuits, where MEMS may be employed to tune the optical properties of devices, as well as for MEMS sensors, where optomechanical devices may be used for shot-noise-limited read-out and back-action modification of the mechanical response.  Moreover, an integrated MEMS-optomechanics technology could allow for up-conversion of low-frequency electrical signals to an optical carrier, mediated by an intermediate mechanical transducer. The ultimate goal of such a conversion scheme would see the realization of coherent, quantum frequency translation between an optical and microwave cavity which shares the same mechanical resonator~\cite{Stannigel2010,Safavi2011,Regal2011,Barzanjeh2011,Wang2012b}.  Different approaches to realize such frequency translation include the use of a silicon nitride membrane vertically stacked within an electronic and optical cavity~\cite{Bagci2014,Andrews2014}, and the creation of piezoelectric nanobeam optomechanical crystals~\cite{Bochmann2013}. 



Our approach to integrating electromechanical and optomechanical devices utilizes a silicon (Si) optomechanical photonic crystal whose mechanical degree of freedom is shared with an electrical capacitor~\cite{Winger2011}. Using a nanoslotted planar photonic crystal slab, it is possible to localize the optical field to the center slot of the slab and the capacitor to the other outer edges of the slab, with both optical and elecrostatic fields connected to the same slab motion.  This avoids losses in the optical path, and in the case of superconducting circuits, avoids optically-induced electrical losses from the breaking of Cooper pairs.  In this kind of planar device, the capacitive element is almost one dimensional, being formed by two parallel metal wires. A strong coupling between the mechanical motion and the optical or electrical mode (in what follows we refer to the lower frequency mode as the electrical mode) can be realized by making the electromagnetic mode volume commensurate with the acoustic wavelength of the mechanical resonance. In the case of near-infrared optics and GHz mechanical resonances, one finds there is a common wavelength scale, whereas for radio-frequency or microwave electrical modes, the mechanical and electrical length scales are vastly different.  In the microwave frequency range this requires decreasing the capacitor gap size to tens of nanometers, where metallic or superconducting boundaries are used to effectively "squeeze" the electric field into a small volume~\cite{Cicak2009a,Sulkko2010}.

In this Article, we push the fabrication limits of the silicon-on-insulator (SOI) cavity-electro-optomechanics platform first presented in Ref.~\cite{Winger2011}, to achieve large electromechanical coupling by engineering a narrow electrode gap while retaining a high optical $Q$-factor.  To this end, we use an \emph{inverse} shadow-mask evaporation which allows us to fabricate electrode gaps as small as $d\approx 30$~nm. Compared to other fabrication techniques, such as focused ion beam milling~\cite{Sulkko2010}, this method maintains a pristine semiconductor surface, avoiding damage to the optical resonator~\cite{Tian2009}. An outline of the paper is as follows.  We begin with a review of the optical, electrical, and mechanical design of the structure, followed by a description of the methods used for device fabrication.  Optical and mechanical characterization of the device are then presented. This is followed by a characterization of both the linear and nonlinear electromechanical coupling.  We conclude by discussing the potential application of these devices for efficient and noise-free microwave-to-optical signal conversion.

\begin{figure}[t!]
\centering
\includegraphics[width=\columnwidth]{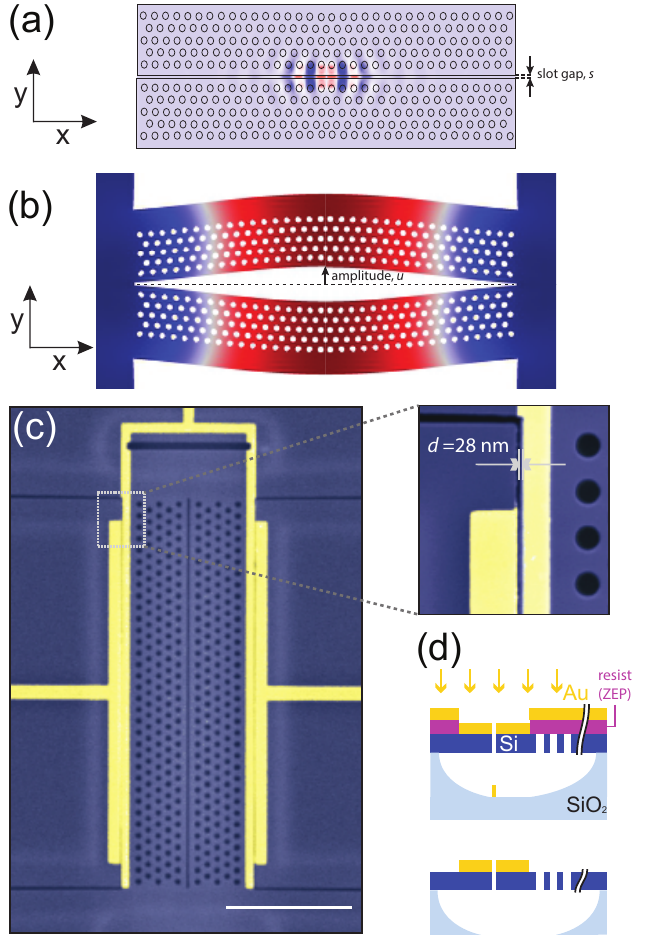}
\caption{(a) Simulated $E_\textrm{y}(\vecb{r})$ electric field component of the optical mode. (b) Total deformation $|\vecb{Q}(\vecb{r})|$ of the fundamental planar flexural mode of the Si slab. (c) Scanning electron microscope (SEM) image of the top-view of a typical realized device. The scale bar is $5$~$\mu$m.  The zoomed-in view to the right depicts the electrode gap, whose size is $d<30$~nm. (d) Illustration of the inverse shadow mask technique.  A mask is created with ZEP resist to define the electrical contacts, while the capacitor electrodes are defined using the etched Si edges. After lift-off and full release, the metal within the gap is removed together with the SiO$_2$ buffer layer.}
\label{fig:1}
\end{figure}

As shown in Fig.~\ref{fig:1}, the electro-optomechanical device studied here is based around a silicon thin-film photonic crystal in which a linear waveguide is formed around a central nanoscale air slot (a so-called W1 slotted waveguide)~\cite{Winger2011,Safavi2010}.  An optical resonant cavity is formed from the waveguide by creating a defect along the axial length of the waveguide in which the parameters of the waveguide are slowly modified. This results in an optical mode confined in the $s \sim 80$~nm air slot and localized to the defect region, as shown in the finite-element-method (FEM) simulation of Fig.~\ref{fig:1}(a).  Mechanical motion of the structure is allowed by undercutting, and suspending the Si device layer. Two additional gaps are fabricated on the outer edge of the two photonic crystal slabs to accommodate capacitor electrodes which connect the mechanical motion of the slabs to an electrical circuit (see Fig.~\ref{fig:1}c). The whole slab structure is clamped on the ends to the underlying SiO$_2$ (BOX) layer and Si substrate, resulting in a fundamental in-plane mode with simulated frequency of $\omegamO/2\pi=67$~MHz. The deformation profile, $|\vecb{Q}(\vecb{r})|$, of the differential motion of the two slabs is shown in Fig.~\ref{fig:1}(b).
A spatial separation between light and metals, as in our device, has a twofold benefit.  From the photonic side it makes metallic-related losses negligible, and from the electrical side it avoids any stray light on the electrodes, which can be crucial if superconducting materials are used.

The main difference between the device studied here and the one reported in Ref.~\cite{Winger2011} resides in the electrode gap size, which has been reduced from $250$~nm to less than $30$~nm using an inverted shadow-mask evaporation technique in a two-layer lithography process (see Fig.~\ref{fig:1}d). In the first step of the device fabrication, a ZEP resist mask is defined with electron beam lithography. The $220$~nm Si device layer of the SOI wafer used in this work is dry etched with an inductively-coupled SF$_6$/C$_4$F$_8$ plasma reactive-ion etching process. After a H$_2$SO$_4$/H$_2$O$_2$ Piranha clean, the device layer is partially undercut in a $15$~second hydrofluoric acid (HF) wet etching step.  The electrical circuit is then defined in a ZEP lift-off process using an aligned electron beam lithography step, followed by deposition and of chromium ($5$~nm) and gold ($50$~nm) layers and then lift-off.  In this last step, the mask for the capacitor electrodes is defined using the partially underetched silicon slot, resulting in a metal gap size defined by the etched slot itself. After lift-off in ZDMAC, the entire optomechanical device is released in HF, whereby the metal deposited between the capacitor electrodes is removed along with the underlying BOX layer. A SEM of the final device is shown in Fig.~\ref{fig:1}(c). As can be seen in the enlarged view, our fabrication technique enables the realization of clean and narrow vacuum electrode gaps of width $d<30$~nm.

\begin{figure}[t!]
\centering
\includegraphics[width=\columnwidth]{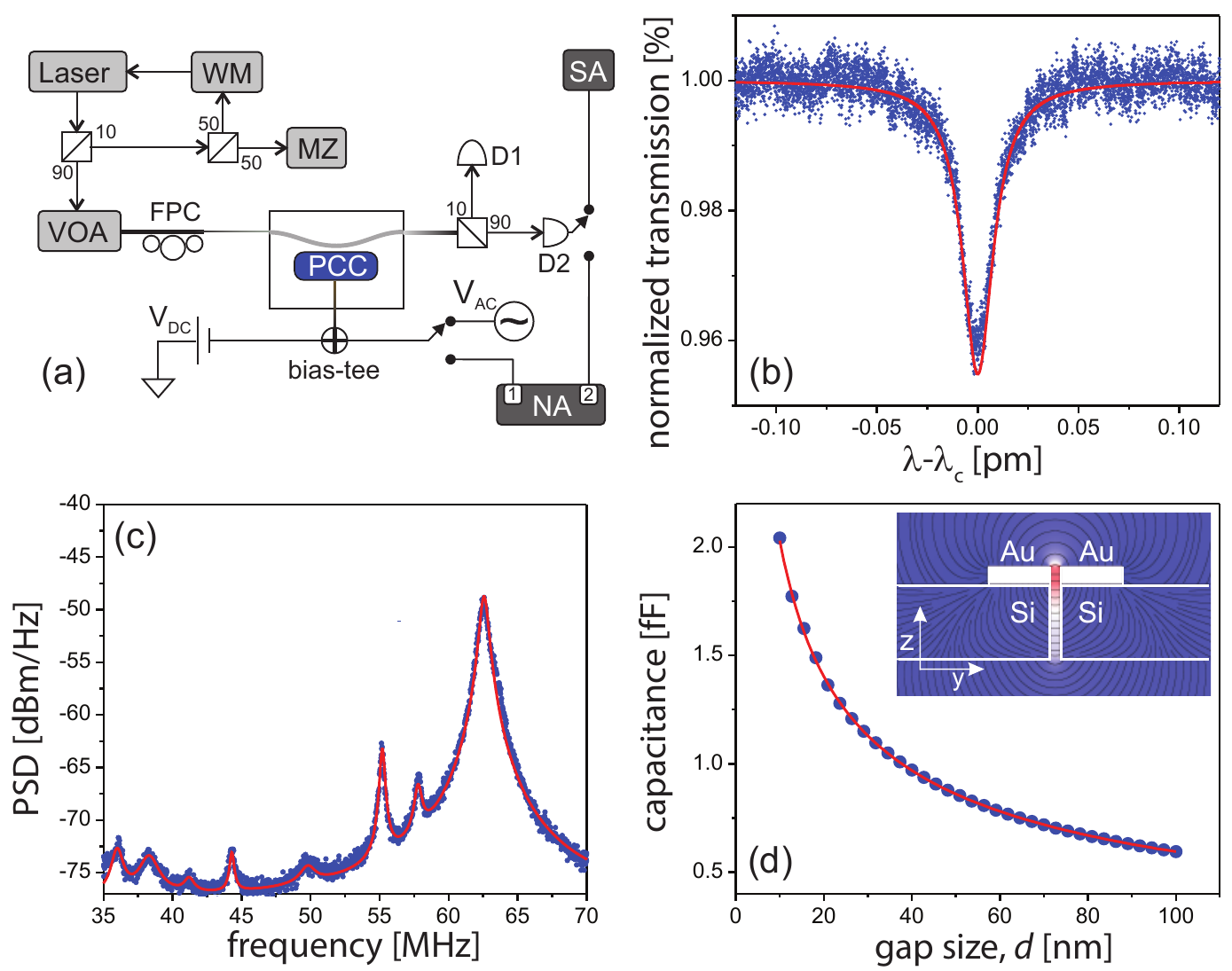}
\caption{(a) Experimental setup for sample characterization. The tunable diode laser is calibrated with a high-finesse Mach Zender interferometer (MZ) and can be locked using a wavemeter (WM) for feedback. After being attenuated (VOA) and polarization controlled (FPC), the evanescent field of a tapered optical fiber is used to couple light into the photonic crystal cavity (PCC). The optical transmission is then measured with a low bandwidth photodetector (D1) in the case of optical spectroscopy, or via a high-speed photodetector (D2) connected to a Spectrum Analyzer (SA) for analysis of the mechanical spectrum. In addition, voltage sources ($\VDC$), a signal generator ($\VAC$) or a Network analyzer (NA) were used to measure the electric response of the device. (b) Measured normalized optical transmission through the photonic crystal of device A. The red line is the best Lorentzian fit of the resonance, showing an intrinsic optical $Q$-factor of $\QoO \sim 9\times 10^4$. (c) Power spectral density of the optically transduced thermal Brownian motion of the structure of device A. The strong resonance peak at $\omegamO/2\pi = 63$~MHz is the fundamental in-plane mechanical mode of the slab structure, whereas the smaller peaks represent weakly coupled (predominantly) out-of-plane modes of the slab. The red line is a multi-Lorentzian fit to the transduced mechanical spectrum. (d) Simulated capacitance as a function of gap size, $d$, for the capacitor on one side of the slab structure. The data fits with a power law (red line) $\Cm = [6.933/d^{0.533}]$~fF, with $d$ in units of nanometers. Inset: simulated $E_\textrm{y}$ electrostatic field amplitude within the capacitor. The field isolines are shown in black.}
\label{fig:2}
\end{figure}

The experimental set-up used to probe the optical, mechanical, and electrical properties of the fabricated device is shown in Fig.~\ref{fig:2}(a).  The device fundamental optical resonance is probed through optical transmission measurements using a tunable external cavity semiconductor diode laser (Newfocus Velocity series) whose frequency tuning is calibrated with an unbalanced fiber Mach-Zender interferometer. Optical coupling to a given device is achieved via a tapered and dimpled optical fiber probe, which when placed in near-field of the photonic crystal cavity allows for evanescent coupling of light between the fiber and cavity~\cite{Michael2007}.  An optical transmission scan, shown in Fig.~\ref{fig:2}(b) for a typical device (device A in what follows), shows a fundamental optical resonance with center wavelength $\lambdacO = 1522.94$~nm and intrinsic quality factor $\QoO = 8.9 \times 10^4$.

The mechanical mode studied in this work is not the fundamental in-plane differential slab mode, but rather the fundamental in-plane mode of only a single slab.  This is an artifact of the measurement technique we employ, in which to mechanically stabilize the optical fiber taper we place it in direct contact with one of the photonic crystal slabs.  This effectively decouples the one slab from the other, and thus we only measure and actuate the motion of the free slab without the taper on it.  The fundamental in-plane mode of the free slab still modifies the air slot gap size in the center of the photonic crystal, and thus induces a frequency shift of the optical cavity resonance which we quantify by an optomechanical coupling parameter $\gOMO$ (defined below) in units of GHz/nm. The transmitted optical power for a laser, frequency locked on the side of the optical cavity resonance, carries a signal corresponding to the thermal Brownian motion of the structure, as shown in Fig.~\ref{fig:2}(c) for device A.  The measured fundamental mechanical mode of this device is centered at $\omegamO/2\pi = 63$~MHz with a corresponding quality factor of $\QmO = 150$, limited by atmospheric pressure squeeze-film damping~\cite{Verbridge2008}. The smaller peaks in the optically-transduced mechanical spectrum are (predominantly) out-of-plane slab modes which are weakly coupled to the optical cavity resonance.

The optomechanical coupling parameter, $\gOMO$, is defined as the fractional change in the energy stored in the optical resonance per unit displacement of the mechanical resonance.  The fractional energy shift ($\dUe$) can be numerically calculated with a perturbative approach, evaluating an integral of the electric and displacement fields (assuming no change in the magnetic energy) of the optical resonance over the dielectric boundaries of the structure~\cite{Johnson2002,Eichenfield2009}:
\begin{equation}\label{eq:pert_int}
\dUe =\frac{\int_{\partial V}{(\tilde{\vecb{Q}}(\vecb{r})\cdot\vec{n})(\Delta\epsilon |\vecb{E}_{\parallel}|^2-\Delta\epsilon^{-1} |\vecb{D}_\perp|^2)d^2r}}{\int_V{\epsilon|\vecb{E}|^2d^3r}},
\end{equation}
where $\epsilon$ is the dielectric constant, $\Delta\epsilon \equiv (\epsilon_1 - \epsilon_2)$ is the difference in the dielectric constant across the boundary between material 1 and material 2, $\Delta\epsilon^{-1} \equiv (1/\epsilon_1 - 1/\epsilon_2)$, and $\vecb{E}_\parallel$ ($\vecb{D}_\perp$) is the parallel (perpendicular) component with respect to the boundary, $\partial V$, of the electric (displacement) field. Here, a generalized coordinate for the mechanical resonance of $u=\textrm{max}(|\vecb{Q(\vecb{r})}|)$ is chosen, corresponding to a normalized displacement field of $\tilde{\vecb{Q}} = \vecb{Q}/\textrm{max}(|\vecb{Q}|)$ in eq.~(\ref{eq:pert_int}). The optomechanical coupling, representing the optical resonance frequency shift per unit displacement amplitude $u$ of the mechanical resonance, is given by $\gOMO = (1/2)\omegacO\dUe$, where $\omegacO$ is the optical resonance frequency and the factor of $1/2$ accounts for the energy in the magnetic field which is decoupled from the mechanical motion.  For the fundamental optical mode coupled to the fundamental in-plane mechanical mode of a single slab, the optomechanical coupling is evaluated to be $\gOMO/2\pi = 36$~GHz/nm.  This value is in good agreement with the experimentally measured value of $37 \pm 6$~GHz/nm for device A, determined from the radiation pressure induced mechanical frequency shift (see for example Ref.~\cite{Winger2011}).  Note that we have implicitly chosen a positive amplitude $u$ to correspond to outward motion of the photonic crystal slabs as shown in Fig.~\ref{fig:1}(b), resulting in a reduced capacitor gap, $d$, and an increased central slot width, $s$.  In what follows we continue to use the same generalized mechanical coordinate, $u=\textrm{max}(|\textbf{Q}|)$, in order to be self-consistent with the calculated value of $\gOMO$.

The motion of the fundamental in-plane mechanical mode can also be detected and actuated via the capacitor on the outer edge of the photonic crystal slab.  The simulated capacitance (for a single side of the slab) is found to be in the range $0.5$-$2$~fF, scaling with the gap $d$ as shown in Fig.~\ref{fig:2}(d).  In analogy with the optomechanical coupling, it is possible to quantify the capacitive, linear electromechanical interaction through the coupling parameter $\GLE \equiv (1/\Cm)(\partial \Cm/\partial u)$~\cite{Bagci2014}, where $\Cm$ is the capacitance modulated by the mechanical mode and $u$ is again the chosen generalized coordinate of the mechanical displacement.  $\GLE$ can be evaluated starting from the electrostatic energy in a capacitor biased with a fixed charge $Q$, $\Ues = Q^2/2\Cm$.  For fixed charge, the fractional electrostatic energy shift due to motion of the mechanical resonance can be numerically calculated for an arbitrary mechanical displacement profile from the unperturbed fields in the capacitor using the same integral as for the calculation of the optomechanical coupling in eq.~(\ref{eq:pert_int}), with $\GLE=-\dUe$.  Here we assume perfectly conducting boundary conditions and zero fields within the metal wires, a good approximation at microwave frequencies and below.  Typically, one would like to couple the mechanical motion to an electrical resonant circuit.  In such a case, the coupling capacitance is closed by an inductor ($L$) to form an LC resonant circuit of frequency $\omegacE/2\pi = (LC)^{-1/2}$.  Assuming capacitive coupling only to the mechanical resonator, the corresponding electromechanical coupling is given by, $\gOME = (-\omegacE/2)\GLE$, in direct analogy to the optomechanical coupling.  Addition of the inductor usually accompanies an unwanted parasitic capacitance, $\Cs$, which is not coupled to the mechanical resonator and which reduces $\GLE$ by a participation factor $\etaE = \Cm/(\Cm+\Cs)$.  Here we will concern ourselves primarily with the coupling parameter $\GLE$, however, in the conclusion we will further discuss coupling to a microwave LC resonator.

\begin{figure}[t!]
\centering
\includegraphics[width=\columnwidth]{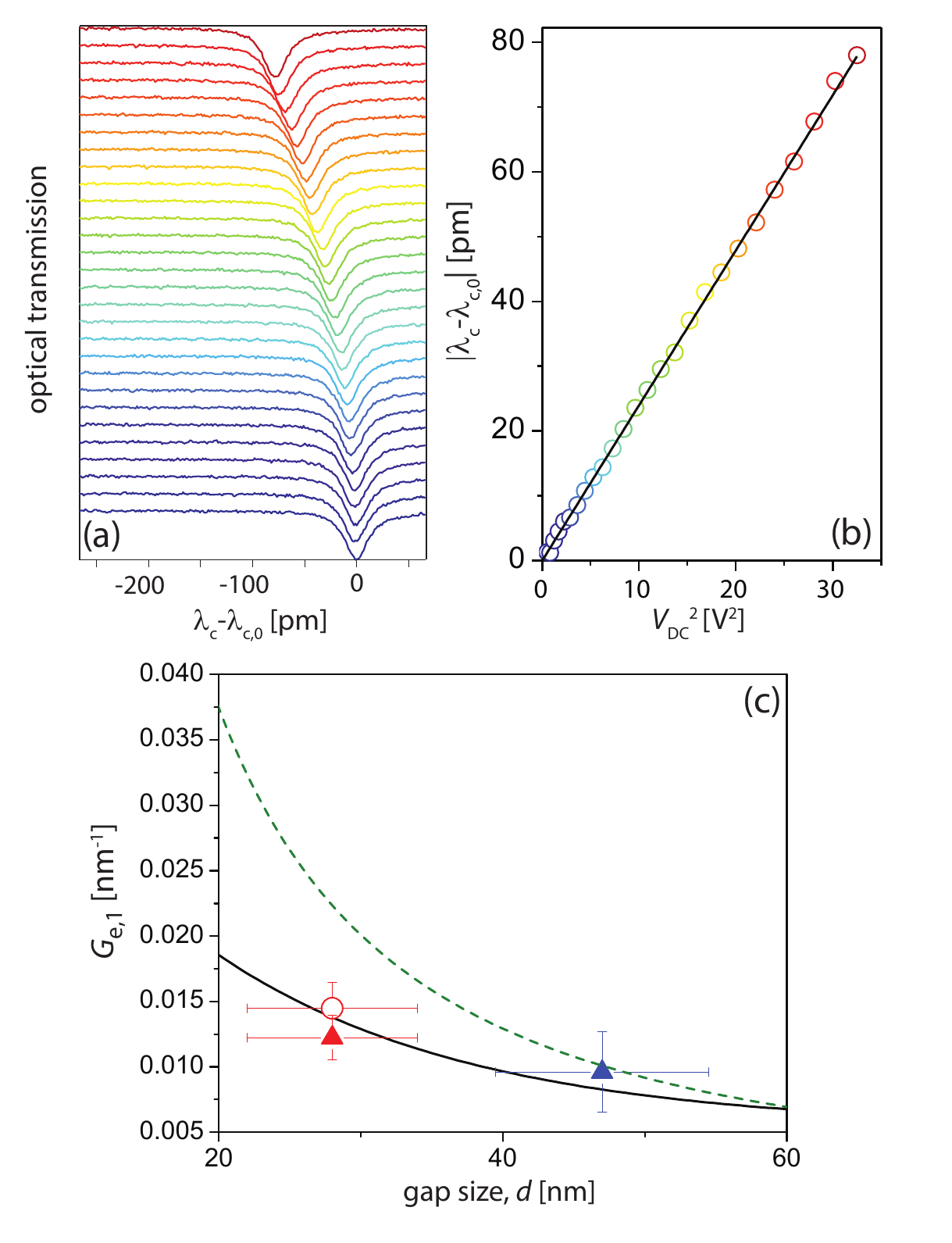}
\caption{(a) Waterfall plot of the optical transmission spectra for the fundamental optical cavity resonance as a function of DC voltage ($\VDC = 0$ to $5.6$~V, $0.2$~V steps). The x-axis has been normalized according to the optical center wavelength at zero applied bias, $\lambda_{c,0}$. (b) Measured optical resonance wavelength (open circles) versus $\VDC^2$.  Solid black curve corresponds to a linear fit to the measured data. (c) Linear coupling parameter extracted from different experiments (DC (circles), AC (triangles)) and for different samples (A (red) and B (blue)).  The solid black line is the numerically simulated linear electromechanical coupling using eq.~(\ref{eq:pert_int}).  The dashed green line is the linear electromechanical coupling calculated using the functional dependence of the capacitance versus gap $d$ shown in Fig.~\ref{fig:2}(d).}
\label{fig:3}
\end{figure}

To measure the linear electromechanical coupling parameter we apply a voltage to the electrodes and measure the resulting mechanical displacement using optical read-out.  This is done for both a static voltage ($\VDC$) and for a small modulated voltage ($\VAC$) at half the resonance frequency of the fundamental in-plane differential mode.  For example, Fig.~\ref{fig:3}(a) shows a waterfall plot of the transmission through the optical cavity for an applied bias voltage of $\VDC=0$ to $5.6$~V for device A. In order to determine the electromechanical coupling from this tuning data, we first consider the force exerted on the fundamental in-plane mechanical mode for a potential difference $V$ across the capacitor,
\begin{equation}\label{eq:F}
\Fcap = \frac{1}{2} \Cm V^2 \GLE,
\end{equation}
where again $\GLE$ is specific to the amplitude coordinate $u$ of a given mechanical resonance.  For a static voltage, the resulting mechanical displacement amplitude $u$ due to $\Fcap$ is inversely proportional to the mode effective spring constant, $\kO=\mO \omegamO^2$.  Here, $\mO$ ($= 10.4$~pg) is the effective motional mass of the fundamental in-plane mechanical mode of a single slab defined as:
\begin{equation}
\mO=\frac{\int{\vecb{Q}^{*}(\vecb{r})\rho(\vecb{r})\vecb{Q}(\vecb{r}) d^3r}}{\text{max}(|\vecb{Q}|^2)},
\end{equation}
where $\rho$ is the mass density of Si. This definition of $\mO$ is consistent with our choice of definition of $u$ corresponding to the maximum amplitude of the mechanical displacement profile.  The effective spring constant, obtained by combining the simulated $\mO$ and the measured mechanical frequency, is equal to $\kO=1.8$~kN/m for device A. This agrees within 18$\%$ of the full numerical simulation of the deformation under a constant load applied to the center point of the mechanical mode. The mechanical deformation shifts the optical resonance frequency according to,
\begin{equation}\label{eq:tunab}
\Delta \omega_\textrm{c,DC} (V)= \frac{\gOMO \Cm \GLE}{2 \kO} \VDC^2 = \alpha \VDC^2,
\end{equation}  
where $\alpha \equiv (\gOMO \Cm \GLE)/(2 \kO)$ is the optical tunability of the structure.

Alternatively, if a modulated voltage $\VAC = V_0\cos(\omega_\textrm{AC} t)$ is applied to the capacitor, the resulting mechanical displacement is filtered by the mechanical response function, with the maximum displacement being enhanced by the mechanical $Q$-factor for an on-resonance capacitive force, $\omega_\textrm{AC} = \omega_\textrm{m}/2$. In this case, the time-average of the optical transmission spectrum assumes a double-dip lineshape (see Ref.~\cite{Winger2011}) with a separation between the minima given by,
\begin{equation}\label{eq:tunab_ac}
\Delta \omega_\textrm{c,AC} = \alpha \QmO V_0^2.
\end{equation}  

Using eqs.~(\ref{eq:tunab}) and (\ref{eq:tunab_ac}) we can extract $\GLE$ from the experimental static and modulated voltage tuning curves, respectively.  As expected from eq.~(\ref{eq:tunab}), the applied static voltage blue shifts the optical resonance as shown in Fig.~\ref{fig:3}(a) for device A with capacitor gap (from SEM images) of $d=28$~nm.  The optical resonance wavelength versus $\VDC^2$ is shown in Fig.~\ref{fig:3}(b) using the same color scale, from which a linear fit yields an optical tunability of $\alpha/2\pi = 0.31$~GHz/V$^2$.  AC measurement of $\alpha$ for device A at a drive frequency of $\omega_\textrm{AC} = \omega_\textrm{m}/2$ yields a small ($20\%$) difference compared to the static voltage measurement, attributable to the impedance mismatch in our AC drive circuit.  A second device (device B) with capacitor gap $d=46$~nm was also measured using the AC tuning method.  Using the measured optomechanical coupling ($\gOMO$) along with the simulated capacitance ($\Cm$) and motional mass ($\mO$), we show a plot of the inferred linear electromechanical coupling parameter $\GLE$ from DC and AC tuning measurements in Fig.~\ref{fig:3}(c) for both device A and device B.  The measured linear electromechanical coupling agrees well with the numerically simulated curve using eq.~(\ref{eq:pert_int}) (solid black curve).


The small capacitor gaps in our devices makes nonlinear terms in the electromechanical interaction relevant.  Electromechanical nonlinearities can be used to generate squeezing~\cite{Rugar1991,Almog2007,Poot2013} and mechanical parametric amplification~\cite{Wu2011,Szorkovszky2014}, as well as for logic operation in the classical~\cite{Mahboob2008} and quantum regime~\cite{Rips2013}.  Assuming the voltage across the capacitor electrodes follows the applied voltage during the mechanical motion of the electrodes~\footnote{This assumption requires that the drive circuit be able to provide the current necessary to the electrodes such that as the electrodes undergo mechanical motion the capacitor voltage follows the applied voltage.  This feedback from the drive circuit modifies the electrostatic force curve similar to the optical feedback in cavity-optomechanics, resulting in a dynamic spring effect.  A wholly different force curve and nonlinear coupling parameter results if we assume a closed capacitor system with fixed charge on the electrodes.}, and expanding the capacitive force to linear order in amplitude $u$, we can define the second order nonlinear coupling parameter as $\GNLE=(1/\Cm)(\partial^2\Cm/\partial u^2)$.  The effective spring constant of the mechanical system is $k_\textrm{tot}=\kO+(1/2)\Cm V^2 \GNLE$~\cite{Rugar1991,Unterreithmeier2009}, resulting in an electrical shift of the \emph{mechanical} resonance frequency given by,
\begin{equation}\label{eq:soft}
\Delta\omegamO=\left(\frac{\omegamO \Cm \GNLE }{4\kO}\right) V^2=\beta V^2,
\end{equation}
where $\beta \equiv (\omegamO \Cm \GNLE)/(4\kO)$ is defined as the \emph{mechanical} tunability. From the mechanical frequency shift versus applied voltage, we can obtain the nonlinear coupling parameter $\GNLE$ in a similar fashion to that used to determine the linear parameter $\GLE$.




Before proceeding to measurements of the nonlinear electromechanical coupling, it is instructive to consider again the capacitor force expression.  If a voltage signal with mixed AC and DC components is fed to the capacitor an additional resonant term appears: 
\begin{equation}
\Fcap \propto V^2 = \VDC^2 + \VAC^2 + 2 \VDC\VAC.
\end{equation}
The third, mixed term, explicitly given by $2V_\textrm{DC} V_0 \cos(\omega_\textrm{AC} t)$, has a maximum mechanical response at $\omega_\textrm{AC}=\omegamO$ and can be suppressed or enhanced by controlling the DC bias. This is a useful and well known property, which allows control of electromechanical nonlinearities~\cite{Agarwal2006,Kaajakari2004}.  The resonant nature of this mixed term is also useful to perform homodyne detection of the coherent mechanical oscillations induced by an AC drive. To this end, we use a network analyzer (NA) of which port $1$ is connected to the capacitor electrodes while port $2$ is connected to the optical photodetector used to read-out the mechanical motion (see Fig.~\ref{fig:2}(a)). The S-parameter $S_{21}$ in such a scheme will therefore carry the amplitude and phase response of the mechanical resonator to the electrical driving force.  

\begin{figure}[t!]
\centering
\includegraphics[width=\columnwidth]{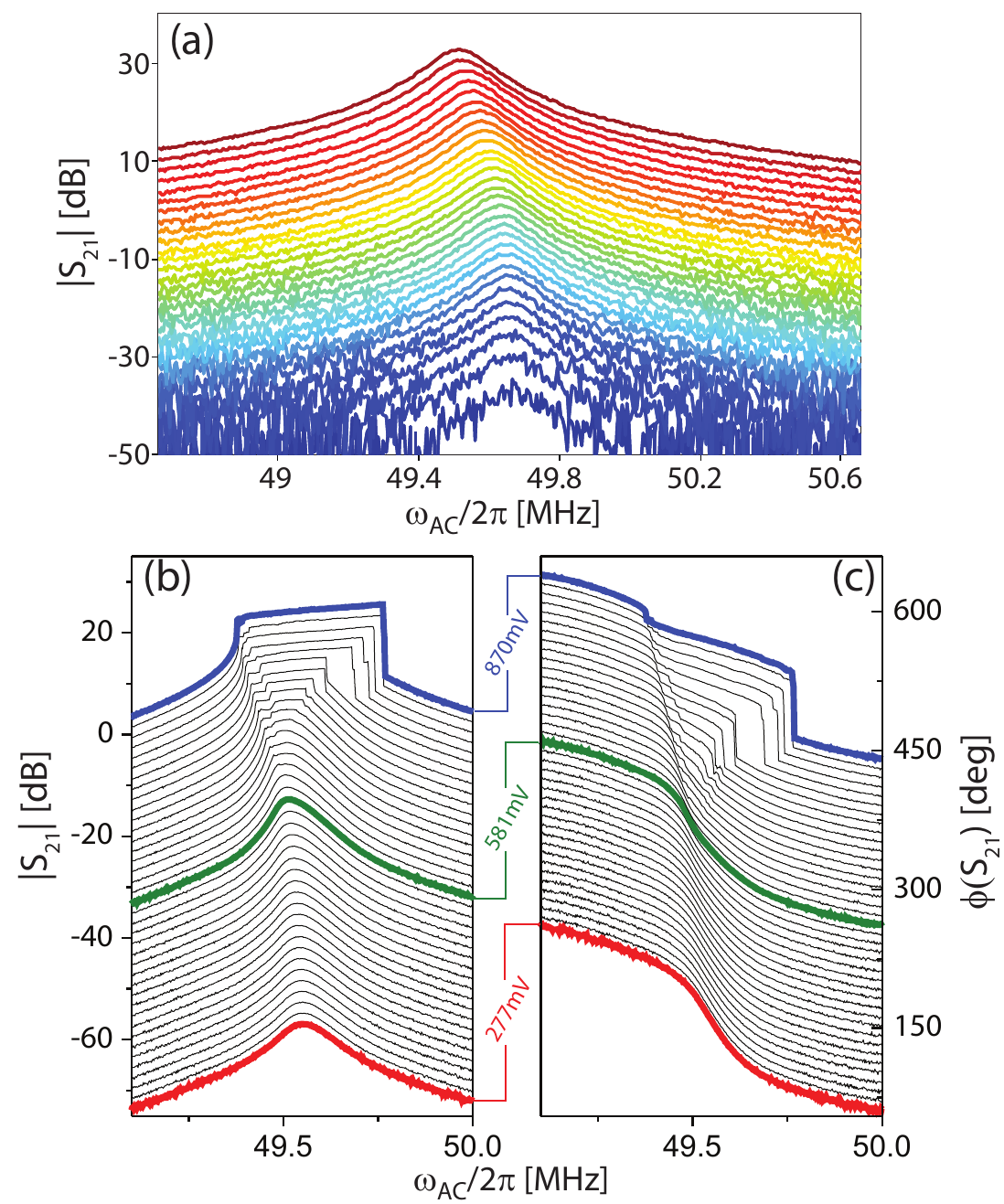}
\caption{Nonlinear electromechanical response of device B.  (a) Measured amplitude response ($|S_{21}|$) for a weak AC voltage probe ($V_0<$1~m\Vrms) and DC bias ranging from $0.2$~V (blue curve) to $6$~V (red curve) in $0.2$~V steps.  Successive spectra are offset from the $\VDC=0.2$~V spectrum in $1$~dB increments.  (b) Measured amplitude response ($|S_{21}|$) of the optically transduced mechanical motion to a strong near-resonant electric drive ($\omega_{\text{AC}} \sim \omegamO$) with fixed $\VDC$=6~V and $V_0$ ranging from $277$-$870$~m\Vrms. Increasing the drive amplitude, the mechanical Lorentzian peak (red) starts to assume the form of a Duffing oscillator (green).  After an instability threshold, the mechanics enters a frequency entrainment regime in which it is completely in phase with the AC drive (blue). (c) Measured relative phase response ($\phi(S_{21})$) of the optically transduced mechanical motion for the same drive conditions as in (b). Note that the phase slope is about the same inside and outside the frequency entrainment region indicating that the phase of the mechanical response is locked to the phase of the drive.  Successive spectra in (b) and (c) are offset from the $\VAC=277$~m\Vrms spectrum in $1$~dB increments.}
\label{fig:4}
\end{figure}

Measurements of the nonlinear electromechanical response were performed on device B, whose fundamental in-plane mechanical mode has a resonance frequency of $\omegamO/2\pi \approx 49$~MHz, intrinsic mechanical $Q$-factor $\QmO = 344$, and spring constant $\kO=1.09$~kN/m.  In order to avoid spurious transduction of large amplitude mechanical motion in these measurements, a fixed probe laser wavelength detuned $\sim 50$~nm from the optical resonance of device B is used. Measurement of the electrostatic modification to the mechanical frequency is first measured, using a weak AC drive voltage of $V_0<$1~m\Vrms~and DC voltages varying from $0.2$ to $6$~V.  The frequency of $\VAC$ is swept using port 1 of the NA and the photodetected signal of the mechanical response is measured on port 2.  For these drive levels ($\VDC \gg V_0$), the mechanical frequency shift in eq.~(\ref{eq:soft}) is dominated by the $\VDC^2$ term.  As can be seen in the waterfall plot of Fig.~\ref{fig:4}(a), the mechanical resonance frequency of device B red-shifts with applied DC voltage, corresponding to electrostatic softening. The extracted resonance frequency scales quadratically with the applied DC bias, yielding a mechanical tunability parameter of $\beta/2\pi=-3.989$~kHz/V$^2$.  This corresponds to a nonlinear coupling parameter of $\GNLE=-3.86 \times 10^{-4}$~nm$^{-2}$, in reasonable agreement with the numerically simulated value of $-4.3 \times 10^{-4}$~nm$^{-2}$.

Measurements of device B under a strong near resonant AC drive ($V_0=277$ to $870$~m\Vrms) and fixed DC bias of $\VDC=6$~V are shown in Fig.~\ref{fig:4}(b).  With increasing AC drive amplitude the mechanical resonance evolves from a symmetric Lorentzian (red curve), to an asymmetric lineshape characteristic of a Duffing oscillator (green curve), and finally into an instability regime (blue curve) in which the mechanical resonance becomes entrained by the driving tone.  The amplitude in this instability regime becomes roughly constant versus drive frequency, and the phase response assumes a constant slope versus frequency in the entrained region as shown in Fig.~\ref{fig:4}(c). Such instabilities are known to occur for parametric driving at frequencies near $2\omegamO/n$, $n$ being an integer corresponding to the $n$th instability ``tongue''~\cite{Lifshitz2003}.  As we are driving near resonance, the parametric instability in this case would correspond to the second tongue ($n=2$), with a theoretical threshold drive amplitude given by~\cite{Lifshitz2003}:
\begin{equation}\label{eq:thresh}
V_\text{0,th}=\frac{\sqrt{2} \kO}{\QmO^{1/2} \Cm \GNLE \VDC}.
\end{equation}
The estimated parametric instability threshold using the $\GNLE$ value from the mechanical softening measurement, is $V_{\text{0,th}}=39$~\Vrms, far larger than the measured instability threshold of only $V_{\textrm{0,th}} \approx 600$~m\Vrms.  This large discrepancy likely indicates that the influence of the large resonant driving force (which is present in addition to the resonant parametric drive) cannot be ignored, and that other mechanical nonlinearities such as the cubic Duffing term also play a role in the onset of the instability. 

To conclude, we consider application of the demonstrated electro-opto-mechanical device to quantum conversion between electrical and optical signals~\cite{Safavi2011,Regal2011}.  In particular, we consider a system in which the capacitor electrodes ($\Cm \sim 1.2$~fF) of the current device are connected together through a wire inductor, forming a lumped element $LC$ resonator in the microwave frequency range.  For such a resonant circuit, the linear electromechanical coupling can be quantified by $\gzeroE \equiv \gOME \xzpfE$, where $\gOME = (-\omegacE/2)\GLE$ and $\xzpfE = (2\hbar \mO \omegamO)^{-1/2}$ is the zero-point amplitude of the mechanical resonator.  A similar relation exists on the optomechanical side, with $\gzeroO \equiv \gOMO \xzpfE$.  Physically, $\gzeroE$ and $\gzeroO$ represent the shift in the microwave and optical resonance for an amplitude of motion equal to the zero-point amplitude of the mechanical resonator, respectively.  For the optomechanical device studied here, $\xzpfE = 3.48$~fm and $\gzeroO/2\pi = 125$~kHz.  For an $LC$ resonator of frequency $\omegacE = 10$~GHz, compatible with current superconducting quantum circuits~\cite{Girvin2009b}, the naively estimated electromechanical coupling rate would be $\gzeroE/2\pi = 250$~Hz. However, one must also consider the stray capacitance associated with adding a large inductance.  Using a planar rectangular spiral inductor with large inductance per wind~\cite{Teufel2011b}, numerical simulations indicate that a $10$~GHz microwave resonance can be realized with a $54$~nH spiral inductor formed on a Si slab (thickness $220$~nm) with wire widths of $500$~nm and inter-wire spacing of $500$~nm.  Such an inductor fits within a $81$~um square, and has a stray capacitance estimated to be $\Cs = 3.26$~fF.  Here we have included a $3$~$\mu$m vacuum layer, corresponding to the undercutting of the BOX layer, between the Si device layer and a $725$~$\mu$m Si handle wafer.  This reduces the estimated electromechanical coupling by a factor of $\etaE = \Cm/(\Cm+\Cs) = 0.27$, to a value of $\gzeroE/2\pi = 70$~Hz for the device studied here.    

The above analysis should be compared against the aluminum superconducting $LC$ resonators of Ref.~\cite{Teufel2011b}, which employ a vertically-layered vacuum gap capacitor with drumhead mechanical modes around $10$~MHz, and have been used to realize strong electromechanical back-action sufficient to cool the mechanical mode to its quantum ground-state and realize efficient microwave-to-optical signal conversion~\cite{Andrews2014}.  The electromechanical coupling in such devices is in the $\gzeroE/2\pi \sim 200$~Hz range, comparable to the estimated value for the nanoslot devices of this work.  Remaining challenges to realizing efficient microwave to optical signal conversion in planar Si devices aimilar to those studied here, include the attainable mechanical and microwave $Q$-factor at milliKelvin temperatures.  Cryogenic temperatures are required both to reduce the thermal noise in the mechanics, as well as to limit the loss in the microwave circuit.  Efficient optical coupling to similar Si optomechanical devices at milliKelvin temperatures has recently been demonstrated~\cite{Meenehan2014}, as have mechanical $Q$-factors greater than $10^5$ at temperatures of $T \sim 10$~K~\cite{SafaviNaeini2013}.  A far more significant challenge will likely be the realization of low-loss superconducting resonators on the SOI wafer platform, in which a high resistivity ($> 1$~kOhm~cm) Si device layer must be used~\cite{O'Connell2008} and the Si surface must be appropriately passivated~\cite{Shim2013}.   

\begin{acknowledgements}
This work was supported by the DARPA MESO program, the AFOSR Hybrid Nanophotonics MURI, the Institute for Quantum Information and Matter, an NSF Physics Frontiers Center with support of the Gordon and Betty Moore Foundation, and the Kavli Nanoscience Institute at Caltech.  AP gratefully acknowledge funding from EU through Marie Curie Actions, project NEMO (GA 298861).  AT acknowledges partial financial support from the ERC through the advanced grant SoulMan.
\end{acknowledgements}


\begin{thebibliography}{44}
\expandafter\ifx\csname natexlab\endcsname\relax\def\natexlab#1{#1}\fi
\expandafter\ifx\csname bibnamefont\endcsname\relax
  \def\bibnamefont#1{#1}\fi
\expandafter\ifx\csname bibfnamefont\endcsname\relax
  \def\bibfnamefont#1{#1}\fi
\expandafter\ifx\csname citenamefont\endcsname\relax
  \def\citenamefont#1{#1}\fi
\expandafter\ifx\csname url\endcsname\relax
  \def\url#1{\texttt{#1}}\fi
\expandafter\ifx\csname urlprefix\endcsname\relax\def\urlprefix{URL }\fi
\providecommand{\bibinfo}[2]{#2}
\providecommand{\eprint}[2][]{\url{#2}}

\bibitem[{\citenamefont{Yazdi et~al.}(1998)\citenamefont{Yazdi, Avazi, and
  Najafi}}]{Yazdi1998}
\bibinfo{author}{\bibfnamefont{N.}~\bibnamefont{Yazdi}},
  \bibinfo{author}{\bibfnamefont{F.}~\bibnamefont{Avazi}}, \bibnamefont{and}
  \bibinfo{author}{\bibfnamefont{K.}~\bibnamefont{Najafi}},
  \bibinfo{journal}{Proc. IEEE} \textbf{\bibinfo{volume}{86}},
  \bibinfo{pages}{1640} (\bibinfo{year}{1998}).

\bibitem[{\citenamefont{Tajima et~al.}(2003)\citenamefont{Tajima, Nishiguchi,
  Chiba, Morita, Abe, Tanioka, Saito, and Esashi}}]{Tajima2003}
\bibinfo{author}{\bibfnamefont{T.}~\bibnamefont{Tajima}},
  \bibinfo{author}{\bibfnamefont{T.}~\bibnamefont{Nishiguchi}},
  \bibinfo{author}{\bibfnamefont{S.}~\bibnamefont{Chiba}},
  \bibinfo{author}{\bibfnamefont{A.}~\bibnamefont{Morita}},
  \bibinfo{author}{\bibfnamefont{M.}~\bibnamefont{Abe}},
  \bibinfo{author}{\bibfnamefont{K.}~\bibnamefont{Tanioka}},
  \bibinfo{author}{\bibfnamefont{N.}~\bibnamefont{Saito}}, \bibnamefont{and}
  \bibinfo{author}{\bibfnamefont{M.}~\bibnamefont{Esashi}},
  \bibinfo{journal}{Microelectron. Eng.} \textbf{\bibinfo{volume}{67-68}},
  \bibinfo{pages}{508} (\bibinfo{year}{2003}).

\bibitem[{\citenamefont{Eatony and Smith}(1997)}]{Eatony1997}
\bibinfo{author}{\bibfnamefont{W.~P.} \bibnamefont{Eatony}} \bibnamefont{and}
  \bibinfo{author}{\bibfnamefont{J.~H.} \bibnamefont{Smith}},
  \bibinfo{journal}{Smart Mater. Struct.} \textbf{\bibinfo{volume}{6}},
  \bibinfo{pages}{530} (\bibinfo{year}{1997}).

\bibitem[{\citenamefont{Cook-Chennault
  et~al.}(2008)\citenamefont{Cook-Chennault, Thambi, and
  Sastry}}]{Cook-Chennault2008}
\bibinfo{author}{\bibfnamefont{K.~A.} \bibnamefont{Cook-Chennault}},
  \bibinfo{author}{\bibfnamefont{N.}~\bibnamefont{Thambi}}, \bibnamefont{and}
  \bibinfo{author}{\bibfnamefont{A.~M.} \bibnamefont{Sastry}},
  \bibinfo{journal}{Smart Mater. Struct.} \textbf{\bibinfo{volume}{17}},
  \bibinfo{pages}{043001} (\bibinfo{year}{2008}).

\bibitem[{\citenamefont{Hanay et~al.}(2012)\citenamefont{Hanay, Kelber, Naik,
  Chi, Hentz, Bullard, Colinet, Duraffourg, and Roukes}}]{Hanay2012}
\bibinfo{author}{\bibfnamefont{M.~S.} \bibnamefont{Hanay}},
  \bibinfo{author}{\bibfnamefont{S.}~\bibnamefont{Kelber}},
  \bibinfo{author}{\bibfnamefont{A.~K.} \bibnamefont{Naik}},
  \bibinfo{author}{\bibfnamefont{D.}~\bibnamefont{Chi}},
  \bibinfo{author}{\bibfnamefont{S.}~\bibnamefont{Hentz}},
  \bibinfo{author}{\bibfnamefont{E.~C.} \bibnamefont{Bullard}},
  \bibinfo{author}{\bibfnamefont{E.}~\bibnamefont{Colinet}},
  \bibinfo{author}{\bibfnamefont{L.}~\bibnamefont{Duraffourg}},
  \bibnamefont{and} \bibinfo{author}{\bibfnamefont{M.~L.}
  \bibnamefont{Roukes}}, \bibinfo{journal}{Nature Nanotech.}
  \textbf{\bibinfo{volume}{7}}, \bibinfo{pages}{602} (\bibinfo{year}{2012}).

\bibitem[{\citenamefont{Manz et~al.}(1990)\citenamefont{Manz, Graber, and
  Widmer}}]{Manz1990}
\bibinfo{author}{\bibfnamefont{A.}~\bibnamefont{Manz}},
  \bibinfo{author}{\bibfnamefont{N.}~\bibnamefont{Graber}}, \bibnamefont{and}
  \bibinfo{author}{\bibfnamefont{H.~M.} \bibnamefont{Widmer}},
  \bibinfo{journal}{Sensors \& Actuators B: Chemical}
  \textbf{\bibinfo{volume}{1}}, \bibinfo{pages}{244} (\bibinfo{year}{1990}).

\bibitem[{\citenamefont{Eichenfield
  et~al.}(2009{\natexlab{a}})\citenamefont{Eichenfield, Camacho, Chan, Vahala,
  and Painter}}]{Eichenfield2009a}
\bibinfo{author}{\bibfnamefont{M.}~\bibnamefont{Eichenfield}},
  \bibinfo{author}{\bibfnamefont{R.~M.} \bibnamefont{Camacho}},
  \bibinfo{author}{\bibfnamefont{J.}~\bibnamefont{Chan}},
  \bibinfo{author}{\bibfnamefont{K.~J.} \bibnamefont{Vahala}},
  \bibnamefont{and} \bibinfo{author}{\bibfnamefont{O.}~\bibnamefont{Painter}},
  \bibinfo{journal}{Nature} \textbf{\bibinfo{volume}{459}},
  \bibinfo{pages}{550} (\bibinfo{year}{2009}{\natexlab{a}}).

\bibitem[{\citenamefont{Eichenfield
  et~al.}(2009{\natexlab{b}})\citenamefont{Eichenfield, Chan, Camacho, Vahala,
  and Painter}}]{Eichenfield2009}
\bibinfo{author}{\bibfnamefont{M.}~\bibnamefont{Eichenfield}},
  \bibinfo{author}{\bibfnamefont{J.}~\bibnamefont{Chan}},
  \bibinfo{author}{\bibfnamefont{R.~M.} \bibnamefont{Camacho}},
  \bibinfo{author}{\bibfnamefont{K.~J.} \bibnamefont{Vahala}},
  \bibnamefont{and} \bibinfo{author}{\bibfnamefont{O.}~\bibnamefont{Painter}},
  \bibinfo{journal}{Nature} \textbf{\bibinfo{volume}{462}}, \bibinfo{pages}{78}
  (\bibinfo{year}{2009}{\natexlab{b}}), ISSN \bibinfo{issn}{0028-0836},
  \urlprefix\url{http://dx.doi.org/10.1038/nature08524}.

\bibitem[{\citenamefont{Cohen and Meenehan}(2013)}]{Cohen2013}
\bibinfo{author}{\bibfnamefont{J.~D.} \bibnamefont{Cohen}} \bibnamefont{and}
  \bibinfo{author}{\bibfnamefont{S.~M.} \bibnamefont{Meenehan}},
  \bibinfo{journal}{Opt. Expr.} \textbf{\bibinfo{volume}{21}},
  \bibinfo{pages}{11227} (\bibinfo{year}{2013}).

\bibitem[{\citenamefont{Anetsberger et~al.}(2010)\citenamefont{Anetsberger,
  Gavartin, Arcizet, Unterreithmeier, Weig, Gorodetsky, Kotthaus, and
  Kippenberg}}]{Anetsberger2010}
\bibinfo{author}{\bibfnamefont{G.}~\bibnamefont{Anetsberger}},
  \bibinfo{author}{\bibfnamefont{E.}~\bibnamefont{Gavartin}},
  \bibinfo{author}{\bibfnamefont{O.}~\bibnamefont{Arcizet}},
  \bibinfo{author}{\bibfnamefont{Q.~P.} \bibnamefont{Unterreithmeier}},
  \bibinfo{author}{\bibfnamefont{E.~M.} \bibnamefont{Weig}},
  \bibinfo{author}{\bibfnamefont{M.~L.} \bibnamefont{Gorodetsky}},
  \bibinfo{author}{\bibfnamefont{J.~P.} \bibnamefont{Kotthaus}},
  \bibnamefont{and} \bibinfo{author}{\bibfnamefont{T.~J.}
  \bibnamefont{Kippenberg}}, \bibinfo{journal}{Phys. Rev. A}
  \textbf{\bibinfo{volume}{82}}, \bibinfo{pages}{061804}
  (\bibinfo{year}{2010}).

\bibitem[{\citenamefont{Chan et~al.}(2011)\citenamefont{Chan, Alegre,
  Safavi-Naeini, Hill, Krause, Groblacher, Aspelmeyer, and Painter}}]{Chan2011}
\bibinfo{author}{\bibfnamefont{J.}~\bibnamefont{Chan}},
  \bibinfo{author}{\bibfnamefont{T.~P.~M.} \bibnamefont{Alegre}},
  \bibinfo{author}{\bibfnamefont{A.~H.} \bibnamefont{Safavi-Naeini}},
  \bibinfo{author}{\bibfnamefont{J.~T.} \bibnamefont{Hill}},
  \bibinfo{author}{\bibfnamefont{A.}~\bibnamefont{Krause}},
  \bibinfo{author}{\bibfnamefont{S.}~\bibnamefont{Groblacher}},
  \bibinfo{author}{\bibfnamefont{M.}~\bibnamefont{Aspelmeyer}},
  \bibnamefont{and} \bibinfo{author}{\bibfnamefont{O.}~\bibnamefont{Painter}},
  \bibinfo{journal}{Nature} \textbf{\bibinfo{volume}{478}}, \bibinfo{pages}{89}
  (\bibinfo{year}{2011}), ISSN \bibinfo{issn}{0028-0836},
  \urlprefix\url{http://dx.doi.org/10.1038/nature10461}.

\bibitem[{\citenamefont{Stannigel et~al.}(2010)\citenamefont{Stannigel, Rabl,
  S\o{}rensen, Zoller, and Lukin}}]{Stannigel2010}
\bibinfo{author}{\bibfnamefont{K.}~\bibnamefont{Stannigel}},
  \bibinfo{author}{\bibfnamefont{P.}~\bibnamefont{Rabl}},
  \bibinfo{author}{\bibfnamefont{A.~S.} \bibnamefont{S\o{}rensen}},
  \bibinfo{author}{\bibfnamefont{P.}~\bibnamefont{Zoller}}, \bibnamefont{and}
  \bibinfo{author}{\bibfnamefont{M.~D.} \bibnamefont{Lukin}},
  \bibinfo{journal}{Phys. Rev. Lett.} \textbf{\bibinfo{volume}{105}},
  \bibinfo{pages}{220501} (\bibinfo{year}{2010}),
  \urlprefix\url{http://link.aps.org/doi/10.1103/PhysRevLett.105.220501}.

\bibitem[{\citenamefont{Safavi-Naeini and Painter}(2011)}]{Safavi2011}
\bibinfo{author}{\bibfnamefont{A.~H.} \bibnamefont{Safavi-Naeini}}
  \bibnamefont{and} \bibinfo{author}{\bibfnamefont{O.}~\bibnamefont{Painter}},
  \bibinfo{journal}{New Journal of Physics} \textbf{\bibinfo{volume}{13}},
  \bibinfo{pages}{013017} (\bibinfo{year}{2011}),
  \urlprefix\url{http://stacks.iop.org/1367-2630/13/i=1/a=013017}.

\bibitem[{\citenamefont{Regal and Lehnert}(2011)}]{Regal2011}
\bibinfo{author}{\bibfnamefont{C.~A.} \bibnamefont{Regal}} \bibnamefont{and}
  \bibinfo{author}{\bibfnamefont{K.~W.} \bibnamefont{Lehnert}},
  \bibinfo{journal}{Journal of Physics: Conference Series}
  \textbf{\bibinfo{volume}{264}}, \bibinfo{pages}{012025}
  (\bibinfo{year}{2011}),
  \urlprefix\url{http://stacks.iop.org/1742-6596/264/i=1/a=012025}.

\bibitem[{\citenamefont{Barzanjeh et~al.}(2012)\citenamefont{Barzanjeh, Abdi,
  Milburn, Tombesi, and Vitali}}]{Barzanjeh2011}
\bibinfo{author}{\bibfnamefont{S.}~\bibnamefont{Barzanjeh}},
  \bibinfo{author}{\bibfnamefont{M.}~\bibnamefont{Abdi}},
  \bibinfo{author}{\bibfnamefont{G.~J.} \bibnamefont{Milburn}},
  \bibinfo{author}{\bibfnamefont{P.}~\bibnamefont{Tombesi}}, \bibnamefont{and}
  \bibinfo{author}{\bibfnamefont{D.}~\bibnamefont{Vitali}},
  \bibinfo{journal}{Phys. Rev. Lett.} \textbf{\bibinfo{volume}{109}},
  \bibinfo{pages}{130503} (\bibinfo{year}{2012}),
  \urlprefix\url{http://link.aps.org/doi/10.1103/PhysRevLett.109.130503}.

\bibitem[{\citenamefont{Wang and Clerk}(2012)}]{Wang2012b}
\bibinfo{author}{\bibfnamefont{Y.-D.} \bibnamefont{Wang}} \bibnamefont{and}
  \bibinfo{author}{\bibfnamefont{A.~A.} \bibnamefont{Clerk}},
  \bibinfo{journal}{New Journal of Physics} \textbf{\bibinfo{volume}{14}},
  \bibinfo{pages}{105010} (\bibinfo{year}{2012}),
  \urlprefix\url{http://stacks.iop.org/1367-2630/14/i=10/a=105010}.

\bibitem[{\citenamefont{Bagci et~al.}(2014)\citenamefont{Bagci, Simonsen,
  Schmid, Villanueva, Zeuthen, Taylor, Sørensen, Usami, A., and
  Polzik}}]{Bagci2014}
\bibinfo{author}{\bibfnamefont{T.}~\bibnamefont{Bagci}},
  \bibinfo{author}{\bibfnamefont{A.}~\bibnamefont{Simonsen}},
  \bibinfo{author}{\bibfnamefont{S.}~\bibnamefont{Schmid}},
  \bibinfo{author}{\bibfnamefont{L.~G.} \bibnamefont{Villanueva}},
  \bibinfo{author}{\bibfnamefont{J.}~\bibnamefont{Zeuthen},
  \bibfnamefont{E.and~Appel}}, \bibinfo{author}{\bibfnamefont{J.~M.}
  \bibnamefont{Taylor}},
  \bibinfo{author}{\bibfnamefont{A.}~\bibnamefont{Sørensen}},
  \bibinfo{author}{\bibfnamefont{K.}~\bibnamefont{Usami}},
  \bibinfo{author}{\bibfnamefont{S.}~\bibnamefont{A.}}, \bibnamefont{and}
  \bibinfo{author}{\bibfnamefont{E.~S.} \bibnamefont{Polzik}},
  \bibinfo{journal}{Nature} \textbf{\bibinfo{volume}{507}}, \bibinfo{pages}{81}
  (\bibinfo{year}{2014}).

\bibitem[{\citenamefont{Andrews et~al.}(2014)\citenamefont{Andrews, Peterson,
  Purdy, Cicak, Simmonds, Regal, and Lehnert}}]{Andrews2014}
\bibinfo{author}{\bibfnamefont{R.~W.} \bibnamefont{Andrews}},
  \bibinfo{author}{\bibfnamefont{R.~W.} \bibnamefont{Peterson}},
  \bibinfo{author}{\bibfnamefont{T.~P.} \bibnamefont{Purdy}},
  \bibinfo{author}{\bibfnamefont{K.}~\bibnamefont{Cicak}},
  \bibinfo{author}{\bibfnamefont{R.~W.} \bibnamefont{Simmonds}},
  \bibinfo{author}{\bibfnamefont{C.~A.} \bibnamefont{Regal}}, \bibnamefont{and}
  \bibinfo{author}{\bibfnamefont{K.~W.} \bibnamefont{Lehnert}},
  \bibinfo{journal}{Nat. Phys.} \textbf{\bibinfo{volume}{10}},
  \bibinfo{pages}{321 } (\bibinfo{year}{2014}).

\bibitem[{\citenamefont{Bochmann et~al.}(2013)\citenamefont{Bochmann,
  Vainsencher, Awschalom, and Cleland}}]{Bochmann2013}
\bibinfo{author}{\bibfnamefont{J.}~\bibnamefont{Bochmann}},
  \bibinfo{author}{\bibfnamefont{A.}~\bibnamefont{Vainsencher}},
  \bibinfo{author}{\bibfnamefont{D.~D.} \bibnamefont{Awschalom}},
  \bibnamefont{and} \bibinfo{author}{\bibfnamefont{A.~N.}
  \bibnamefont{Cleland}}, \bibinfo{journal}{Nat Phys}
  \textbf{\bibinfo{volume}{9}}, \bibinfo{pages}{712} (\bibinfo{year}{2013}),
  ISSN \bibinfo{issn}{1745-2473},
  \urlprefix\url{http://dx.doi.org/10.1038/nphys2748}.

\bibitem[{\citenamefont{Winger et~al.}(2011)\citenamefont{Winger, Blasius,
  Alegre, Safavi-Naeini, Meenehan, Cohen, Stobbe, and Painter}}]{Winger2011}
\bibinfo{author}{\bibfnamefont{M.}~\bibnamefont{Winger}},
  \bibinfo{author}{\bibfnamefont{T.~D.} \bibnamefont{Blasius}},
  \bibinfo{author}{\bibfnamefont{T.~P.~M.} \bibnamefont{Alegre}},
  \bibinfo{author}{\bibfnamefont{A.~H.} \bibnamefont{Safavi-Naeini}},
  \bibinfo{author}{\bibfnamefont{S.}~\bibnamefont{Meenehan}},
  \bibinfo{author}{\bibfnamefont{J.}~\bibnamefont{Cohen}},
  \bibinfo{author}{\bibfnamefont{S.}~\bibnamefont{Stobbe}}, \bibnamefont{and}
  \bibinfo{author}{\bibfnamefont{O.}~\bibnamefont{Painter}},
  \bibinfo{journal}{Opt. Express} \textbf{\bibinfo{volume}{19}},
  \bibinfo{pages}{24905} (\bibinfo{year}{2011}),
  \urlprefix\url{http://www.opticsexpress.org/abstract.cfm?URI=oe-19-25-24905}.

\bibitem[{\citenamefont{Cicak et~al.}(2009)\citenamefont{Cicak, Allman, Strong,
  Osborn, and Simmonds}}]{Cicak2009a}
\bibinfo{author}{\bibfnamefont{K.}~\bibnamefont{Cicak}},
  \bibinfo{author}{\bibfnamefont{M.}~\bibnamefont{Allman}},
  \bibinfo{author}{\bibfnamefont{J.}~\bibnamefont{Strong}},
  \bibinfo{author}{\bibfnamefont{K.}~\bibnamefont{Osborn}}, \bibnamefont{and}
  \bibinfo{author}{\bibfnamefont{R.}~\bibnamefont{Simmonds}},
  \bibinfo{journal}{Applied Superconductivity, IEEE Transactions on}
  \textbf{\bibinfo{volume}{19}}, \bibinfo{pages}{948 } (\bibinfo{year}{2009}),
  ISSN \bibinfo{issn}{1051-8223}.

\bibitem[{\citenamefont{Sulkko et~al.}(2010)\citenamefont{Sulkko,
  SillanpaÌˆaÌˆ, HaÌˆkkinen, Lechner, Helle, Fefferman, Parpia, and
  Hakonen}}]{Sulkko2010}
\bibinfo{author}{\bibfnamefont{J.}~\bibnamefont{Sulkko}},
  \bibinfo{author}{\bibfnamefont{M.~A.} \bibnamefont{SillanpaÌˆaÌˆ}},
  \bibinfo{author}{\bibfnamefont{P.}~\bibnamefont{HaÌˆkkinen}},
  \bibinfo{author}{\bibfnamefont{L.}~\bibnamefont{Lechner}},
  \bibinfo{author}{\bibfnamefont{M.}~\bibnamefont{Helle}},
  \bibinfo{author}{\bibfnamefont{A.}~\bibnamefont{Fefferman}},
  \bibinfo{author}{\bibfnamefont{J.}~\bibnamefont{Parpia}}, \bibnamefont{and}
  \bibinfo{author}{\bibfnamefont{P.~J.} \bibnamefont{Hakonen}},
  \bibinfo{journal}{Nano Letters} \textbf{\bibinfo{volume}{10}},
  \bibinfo{pages}{4884} (\bibinfo{year}{2010}),
  \eprint{http://pubs.acs.org/doi/pdf/10.1021/nl102771p},
  \urlprefix\url{http://pubs.acs.org/doi/abs/10.1021/nl102771p}.

\bibitem[{\citenamefont{Tian et~al.}(2009)\citenamefont{Tian, Yan, Liu, Luo,
  Zhang, Li, and Qiu}}]{Tian2009}
\bibinfo{author}{\bibfnamefont{J.}~\bibnamefont{Tian}},
  \bibinfo{author}{\bibfnamefont{W.}~\bibnamefont{Yan}},
  \bibinfo{author}{\bibfnamefont{Y.}~\bibnamefont{Liu}},
  \bibinfo{author}{\bibfnamefont{J.}~\bibnamefont{Luo}},
  \bibinfo{author}{\bibfnamefont{D.}~\bibnamefont{Zhang}},
  \bibinfo{author}{\bibfnamefont{Z.}~\bibnamefont{Li}}, \bibnamefont{and}
  \bibinfo{author}{\bibfnamefont{M.}~\bibnamefont{Qiu}}, \bibinfo{journal}{J.
  Lightw. Technol.} \textbf{\bibinfo{volume}{27}}, \bibinfo{pages}{4306}
  (\bibinfo{year}{2009}).

\bibitem[{\citenamefont{Safavi-Naeini et~al.}(2010)\citenamefont{Safavi-Naeini,
  Alegre, Winger, and Painter}}]{Safavi2010}
\bibinfo{author}{\bibfnamefont{A.~H.} \bibnamefont{Safavi-Naeini}},
  \bibinfo{author}{\bibfnamefont{T.~P.~M.} \bibnamefont{Alegre}},
  \bibinfo{author}{\bibfnamefont{M.}~\bibnamefont{Winger}}, \bibnamefont{and}
  \bibinfo{author}{\bibfnamefont{O.}~\bibnamefont{Painter}},
  \bibinfo{journal}{Applied Physics Letters} \textbf{\bibinfo{volume}{97}},
  \bibinfo{eid}{181106} (pages~\bibinfo{numpages}{3}) (\bibinfo{year}{2010}),
  \urlprefix\url{http://link.aip.org/link/?APL/97/181ß106/1}.

\bibitem[{\citenamefont{Michael et~al.}(2007)\citenamefont{Michael, Borselli,
  Johnson, Chrystal, and O.}}]{Michael2007}
\bibinfo{author}{\bibfnamefont{C.~P.} \bibnamefont{Michael}},
  \bibinfo{author}{\bibfnamefont{M.}~\bibnamefont{Borselli}},
  \bibinfo{author}{\bibfnamefont{T.~J.} \bibnamefont{Johnson}},
  \bibinfo{author}{\bibfnamefont{C.}~\bibnamefont{Chrystal}}, \bibnamefont{and}
  \bibinfo{author}{\bibfnamefont{P.}~\bibnamefont{O.}}, \bibinfo{journal}{Opt.
  Expr.} \textbf{\bibinfo{volume}{15}}, \bibinfo{pages}{4745}
  (\bibinfo{year}{2007}).

\bibitem[{\citenamefont{Verbridge et~al.}(2008)\citenamefont{Verbridge,
  Craighead, and Parpia}}]{Verbridge2008}
\bibinfo{author}{\bibfnamefont{S.~S.} \bibnamefont{Verbridge}},
  \bibinfo{author}{\bibfnamefont{H.~G.} \bibnamefont{Craighead}},
  \bibnamefont{and} \bibinfo{author}{\bibfnamefont{J.~M.}
  \bibnamefont{Parpia}}, \bibinfo{journal}{Appl. Phys. Lett.}
  \textbf{\bibinfo{volume}{92}}, \bibinfo{pages}{013112}
  (\bibinfo{year}{2008}).

\bibitem[{\citenamefont{Johnson et~al.}(2002)\citenamefont{Johnson, Ibanescu,
  Skorobogatiy, Weisberg, Joannopoulos, and Fink}}]{Johnson2002}
\bibinfo{author}{\bibfnamefont{S.~G.} \bibnamefont{Johnson}},
  \bibinfo{author}{\bibfnamefont{M.}~\bibnamefont{Ibanescu}},
  \bibinfo{author}{\bibfnamefont{M.~A.} \bibnamefont{Skorobogatiy}},
  \bibinfo{author}{\bibfnamefont{O.}~\bibnamefont{Weisberg}},
  \bibinfo{author}{\bibfnamefont{J.~D.} \bibnamefont{Joannopoulos}},
  \bibnamefont{and} \bibinfo{author}{\bibfnamefont{Y.}~\bibnamefont{Fink}},
  \bibinfo{journal}{Phys. Rev. E} \textbf{\bibinfo{volume}{65}},
  \bibinfo{eid}{066611} (\bibinfo{year}{2002}).

\bibitem[{\citenamefont{Rugar and Gr\"utter}(1991)}]{Rugar1991}
\bibinfo{author}{\bibfnamefont{D.}~\bibnamefont{Rugar}} \bibnamefont{and}
  \bibinfo{author}{\bibfnamefont{P.}~\bibnamefont{Gr\"utter}},
  \bibinfo{journal}{Phys. Rev. Lett.} \textbf{\bibinfo{volume}{67}},
  \bibinfo{pages}{699} (\bibinfo{year}{1991}),
  \urlprefix\url{http://link.aps.org/doi/10.1103/PhysRevLett.67.699}.

\bibitem[{\citenamefont{Almog et~al.}(2007)\citenamefont{Almog, Zaitsev,
  Shtempluck, and Buks}}]{Almog2007}
\bibinfo{author}{\bibfnamefont{R.}~\bibnamefont{Almog}},
  \bibinfo{author}{\bibfnamefont{S.}~\bibnamefont{Zaitsev}},
  \bibinfo{author}{\bibfnamefont{O.}~\bibnamefont{Shtempluck}},
  \bibnamefont{and} \bibinfo{author}{\bibfnamefont{E.}~\bibnamefont{Buks}},
  \bibinfo{journal}{Phys. Rev. Lett.} \textbf{\bibinfo{volume}{98}},
  \bibinfo{pages}{078103} (\bibinfo{year}{2007}),
  \urlprefix\url{http://link.aps.org/doi/10.1103/PhysRevLett.98.078103}.

\bibitem[{\citenamefont{Poot and Tang}(2013)}]{Poot2013}
\bibinfo{author}{\bibfnamefont{M.}~\bibnamefont{Poot}} \bibnamefont{and}
  \bibinfo{author}{\bibfnamefont{H.~X.} \bibnamefont{Tang}}, in
  \emph{\bibinfo{booktitle}{CLEO: 2013}} (\bibinfo{publisher}{Optical Society
  of America}, \bibinfo{year}{2013}), p. \bibinfo{pages}{CW3F.7},
  \urlprefix\url{http://www.opticsinfobase.org/abstract.cfm?URI=CLEO_SI-2013-CW3F.7}.

\bibitem[{\citenamefont{Wu and Zhong}(2011)}]{Wu2011}
\bibinfo{author}{\bibfnamefont{C.~C.} \bibnamefont{Wu}} \bibnamefont{and}
  \bibinfo{author}{\bibfnamefont{Z.}~\bibnamefont{Zhong}},
  \bibinfo{journal}{Nano Letters} \textbf{\bibinfo{volume}{11}},
  \bibinfo{pages}{1448} (\bibinfo{year}{2011}),
  \eprint{http://pubs.acs.org/doi/pdf/10.1021/nl1039549},
  \urlprefix\url{http://pubs.acs.org/doi/abs/10.1021/nl1039549}.

\bibitem[{\citenamefont{Szorkovszky et~al.}(2014)\citenamefont{Szorkovszky,
  Clerk, Doherty, and Bowen}}]{Szorkovszky2014}
\bibinfo{author}{\bibfnamefont{A.}~\bibnamefont{Szorkovszky}},
  \bibinfo{author}{\bibfnamefont{A.~A.} \bibnamefont{Clerk}},
  \bibinfo{author}{\bibfnamefont{A.~C.} \bibnamefont{Doherty}},
  \bibnamefont{and} \bibinfo{author}{\bibfnamefont{W.~P.} \bibnamefont{Bowen}},
  \bibinfo{journal}{New J. Phys.} \textbf{\bibinfo{volume}{16}},
  \bibinfo{pages}{043023} (\bibinfo{year}{2014}).

\bibitem[{\citenamefont{Mahboob and Yamaguchi}(2008)}]{Mahboob2008}
\bibinfo{author}{\bibfnamefont{I.}~\bibnamefont{Mahboob}} \bibnamefont{and}
  \bibinfo{author}{\bibfnamefont{H.}~\bibnamefont{Yamaguchi}},
  \bibinfo{journal}{Nat Nano} \textbf{\bibinfo{volume}{3}},
  \bibinfo{pages}{275} (\bibinfo{year}{2008}), ISSN \bibinfo{issn}{1748-3387},
  \urlprefix\url{http://dx.doi.org/10.1038/nnano.2008.84}.

\bibitem[{\citenamefont{Rips and Hartmann}(2013)}]{Rips2013}
\bibinfo{author}{\bibfnamefont{S.}~\bibnamefont{Rips}} \bibnamefont{and}
  \bibinfo{author}{\bibfnamefont{M.~J.} \bibnamefont{Hartmann}},
  \bibinfo{journal}{Phys. Rev. Lett.} \textbf{\bibinfo{volume}{110}},
  \bibinfo{pages}{120503} (\bibinfo{year}{2013}),
  \urlprefix\url{http://link.aps.org/doi/10.1103/PhysRevLett.110.120503}.

\bibitem[{\citenamefont{Unterreithmeier
  et~al.}(2009)\citenamefont{Unterreithmeier, Weig, and
  Kotthaus}}]{Unterreithmeier2009}
\bibinfo{author}{\bibfnamefont{Q.~P.} \bibnamefont{Unterreithmeier}},
  \bibinfo{author}{\bibfnamefont{E.~M.} \bibnamefont{Weig}}, \bibnamefont{and}
  \bibinfo{author}{\bibfnamefont{J.~P.} \bibnamefont{Kotthaus}},
  \bibinfo{journal}{Nature} \textbf{\bibinfo{volume}{458}},
  \bibinfo{pages}{1001} (\bibinfo{year}{2009}), ISSN \bibinfo{issn}{0028-0836},
  \urlprefix\url{http://dx.doi.org/10.1038/nature07932}.

\bibitem[{\citenamefont{Agarwal et~al.}(2006)\citenamefont{Agarwal, Park,
  Candler, Kim, Hopcroft, Chandorkar, Jha, Melamud, Kenny, and
  Murmann}}]{Agarwal2006}
\bibinfo{author}{\bibfnamefont{M.}~\bibnamefont{Agarwal}},
  \bibinfo{author}{\bibfnamefont{K.-K.} \bibnamefont{Park}},
  \bibinfo{author}{\bibfnamefont{R.}~\bibnamefont{Candler}},
  \bibinfo{author}{\bibfnamefont{B.}~\bibnamefont{Kim}},
  \bibinfo{author}{\bibfnamefont{M.}~\bibnamefont{Hopcroft}},
  \bibinfo{author}{\bibfnamefont{S.~A.} \bibnamefont{Chandorkar}},
  \bibinfo{author}{\bibfnamefont{C.}~\bibnamefont{Jha}},
  \bibinfo{author}{\bibfnamefont{R.}~\bibnamefont{Melamud}},
  \bibinfo{author}{\bibfnamefont{T.}~\bibnamefont{Kenny}}, \bibnamefont{and}
  \bibinfo{author}{\bibfnamefont{B.}~\bibnamefont{Murmann}}, in
  \emph{\bibinfo{booktitle}{International Frequency Control Symposium and
  Exposition, 2006 IEEE}} (\bibinfo{year}{2006}), pp.
  \bibinfo{pages}{209--212}.

\bibitem[{\citenamefont{Kaajakari et~al.}(2005)\citenamefont{Kaajakari,
  Mattila, Lipsanen, and Oja}}]{Kaajakari2004}
\bibinfo{author}{\bibfnamefont{V.}~\bibnamefont{Kaajakari}},
  \bibinfo{author}{\bibfnamefont{T.}~\bibnamefont{Mattila}},
  \bibinfo{author}{\bibfnamefont{A.}~\bibnamefont{Lipsanen}}, \bibnamefont{and}
  \bibinfo{author}{\bibfnamefont{A.}~\bibnamefont{Oja}},
  \bibinfo{journal}{Sensors and Actuators A: Physical}
  \textbf{\bibinfo{volume}{120}}, \bibinfo{pages}{64 } (\bibinfo{year}{2005}),
  ISSN \bibinfo{issn}{0924-4247},
  \urlprefix\url{http://www.sciencedirect.com/science/article/pii/S092442470400826X}.

\bibitem[{\citenamefont{Lifshitz and Cross}(2003)}]{Lifshitz2003}
\bibinfo{author}{\bibfnamefont{R.}~\bibnamefont{Lifshitz}} \bibnamefont{and}
  \bibinfo{author}{\bibfnamefont{M.~C.} \bibnamefont{Cross}},
  \bibinfo{journal}{Phys. Rev. B} \textbf{\bibinfo{volume}{67}},
  \bibinfo{pages}{134302} (\bibinfo{year}{2003}).

\bibitem[{\citenamefont{Girvin et~al.}(2009)\citenamefont{Girvin, Devoret, and
  Schoelkopf}}]{Girvin2009b}
\bibinfo{author}{\bibfnamefont{S.~M.} \bibnamefont{Girvin}},
  \bibinfo{author}{\bibfnamefont{M.~H.} \bibnamefont{Devoret}},
  \bibnamefont{and} \bibinfo{author}{\bibfnamefont{R.~J.}
  \bibnamefont{Schoelkopf}}, \bibinfo{journal}{Phys. Scr.}
  \textbf{\bibinfo{volume}{2009}}, \bibinfo{pages}{014012}
  (\bibinfo{year}{2009}),
  \urlprefix\url{http://stacks.iop.org/1402-4896/2009/i=T137/a=014012}.

\bibitem[{\citenamefont{Teufel et~al.}(2011)\citenamefont{Teufel, Donner, Li,
  Harlow, Allman, Cicak, Sirois, Whittaker, Lehnert, and
  Simmonds}}]{Teufel2011b}
\bibinfo{author}{\bibfnamefont{J.~D.} \bibnamefont{Teufel}},
  \bibinfo{author}{\bibfnamefont{T.}~\bibnamefont{Donner}},
  \bibinfo{author}{\bibfnamefont{D.}~\bibnamefont{Li}},
  \bibinfo{author}{\bibfnamefont{J.~W.} \bibnamefont{Harlow}},
  \bibinfo{author}{\bibfnamefont{M.~S.} \bibnamefont{Allman}},
  \bibinfo{author}{\bibfnamefont{K.}~\bibnamefont{Cicak}},
  \bibinfo{author}{\bibfnamefont{A.~J.} \bibnamefont{Sirois}},
  \bibinfo{author}{\bibfnamefont{J.~D.} \bibnamefont{Whittaker}},
  \bibinfo{author}{\bibfnamefont{K.~W.} \bibnamefont{Lehnert}},
  \bibnamefont{and} \bibinfo{author}{\bibfnamefont{R.~W.}
  \bibnamefont{Simmonds}}, \bibinfo{journal}{Nature}
  \textbf{\bibinfo{volume}{475}}, \bibinfo{pages}{359} (\bibinfo{year}{2011}),
  ISSN \bibinfo{issn}{0028-0836},
  \urlprefix\url{http://dx.doi.org/10.1038/nature10261}.

\bibitem[{\citenamefont{Meenehan et~al.}(2014)\citenamefont{Meenehan, Cohen,
  Groeblacher, Hill, Safavi-Naeini, Aspelmeyer, and Painter}}]{Meenehan2014}
\bibinfo{author}{\bibfnamefont{S.~M.} \bibnamefont{Meenehan}},
  \bibinfo{author}{\bibfnamefont{J.~D.} \bibnamefont{Cohen}},
  \bibinfo{author}{\bibfnamefont{S.}~\bibnamefont{Groeblacher}},
  \bibinfo{author}{\bibfnamefont{J.~T.} \bibnamefont{Hill}},
  \bibinfo{author}{\bibfnamefont{A.~H.} \bibnamefont{Safavi-Naeini}},
  \bibinfo{author}{\bibfnamefont{M.}~\bibnamefont{Aspelmeyer}},
  \bibnamefont{and} \bibinfo{author}{\bibfnamefont{O.}~\bibnamefont{Painter}},
  \bibinfo{journal}{arXiv} \textbf{\bibinfo{volume}{1403.3703}}
  (\bibinfo{year}{2014}).

\bibitem[{\citenamefont{Safavi-Naeini et~al.}(2013)\citenamefont{Safavi-Naeini,
  Groeblacher, Hill, Chan, Aspelmeyer, and Painter}}]{SafaviNaeini2013}
\bibinfo{author}{\bibfnamefont{A.~H.} \bibnamefont{Safavi-Naeini}},
  \bibinfo{author}{\bibfnamefont{S.}~\bibnamefont{Groeblacher}},
  \bibinfo{author}{\bibfnamefont{J.~T.} \bibnamefont{Hill}},
  \bibinfo{author}{\bibfnamefont{J.}~\bibnamefont{Chan}},
  \bibinfo{author}{\bibfnamefont{M.}~\bibnamefont{Aspelmeyer}},
  \bibnamefont{and} \bibinfo{author}{\bibfnamefont{O.}~\bibnamefont{Painter}},
  \bibinfo{journal}{Nature} \textbf{\bibinfo{volume}{500}},
  \bibinfo{pages}{185} (\bibinfo{year}{2013}).

\bibitem[{\citenamefont{O'Connell et~al.}(2008)\citenamefont{O'Connell,
  Ansmann, Bialczak, Hofheinz, Katz, Lucero, McKenney, Neeley, Wang, Weig
  et~al.}}]{O'Connell2008}
\bibinfo{author}{\bibfnamefont{A.~D.} \bibnamefont{O'Connell}},
  \bibinfo{author}{\bibfnamefont{M.}~\bibnamefont{Ansmann}},
  \bibinfo{author}{\bibfnamefont{R.~C.} \bibnamefont{Bialczak}},
  \bibinfo{author}{\bibfnamefont{M.}~\bibnamefont{Hofheinz}},
  \bibinfo{author}{\bibfnamefont{N.}~\bibnamefont{Katz}},
  \bibinfo{author}{\bibfnamefont{E.}~\bibnamefont{Lucero}},
  \bibinfo{author}{\bibfnamefont{C.}~\bibnamefont{McKenney}},
  \bibinfo{author}{\bibfnamefont{M.}~\bibnamefont{Neeley}},
  \bibinfo{author}{\bibfnamefont{H.}~\bibnamefont{Wang}},
  \bibinfo{author}{\bibfnamefont{E.~M.} \bibnamefont{Weig}},
  \bibnamefont{et~al.}, \bibinfo{journal}{Appl. Phys. Lett.}
  \textbf{\bibinfo{volume}{92}}, \bibinfo{pages}{112903}
  (\bibinfo{year}{2008}),
  \urlprefix\url{http://link.aip.org/link/?APL/92/112903/1}.

\bibitem[{\citenamefont{Shim et~al.}(2013)\citenamefont{Shim, Raskin, Neve, and
  Rais-Zadeh}}]{Shim2013}
\bibinfo{author}{\bibfnamefont{Y.}~\bibnamefont{Shim}},
  \bibinfo{author}{\bibfnamefont{J.-P.} \bibnamefont{Raskin}},
  \bibinfo{author}{\bibfnamefont{C.~R.} \bibnamefont{Neve}}, \bibnamefont{and}
  \bibinfo{author}{\bibfnamefont{M.}~\bibnamefont{Rais-Zadeh}},
  \bibinfo{journal}{IEEE Microwave and Wireless Components Letters}
  \textbf{\bibinfo{volume}{23}}, \bibinfo{pages}{632} (\bibinfo{year}{2013}).

\end{thebibliography}

\end{document}